\documentclass[12pt]{article}

\usepackage[centertags]{amsmath}
\usepackage{latexsym,amssymb,color,mathtools}
\usepackage{amsfonts}
\usepackage{newlfont}
\usepackage{epsfig}
\usepackage{graphicx}
\usepackage{setspace}
\usepackage{theorem}
\usepackage{nomencl}
\usepackage{units}
\usepackage{lscape}
\usepackage{rotating}
\usepackage[normalem]{ulem} 
\usepackage{booktabs}

\usepackage{color}

\usepackage{bm}
\usepackage{cite}
\usepackage{ifthen}

\usepackage[plainpages=false,colorlinks=true,citecolor=black,
filecolor=black,linkcolor=blue,urlcolor=blue,backref=page,
pagebackref,bookmarks=true,linktocpage=true]{hyperref}

\renewcommand{\nomgroup}[1]{
  \ifthenelse{\equal{#1}{A}}
    {\item[\emph{Roman Letters}]}
    {\ifthenelse{\equal{#1}{G}}
      {\item[\emph{Greek Letters}]}
      {\ifthenelse{\equal{#1}{U}}
        {\item[\emph{Superscripts}]}
        {\ifthenelse{\equal{#1}{L}}
          {\item[\emph{Subscripts}]}
          {\ifthenelse{\equal{#1}{C}}
            {\item[\emph{Chemical Components}]}
          {}
          }
        }
      }
  }
}

\newcommand{\invisible}[1]{}

\newcommand\RR{{\mathrm{I\kern-.16em R}}}

\def\<{\left< }
\def\>{ \right>}

\def\bi{{\bf i}}
\newcommand{\be}{\begin{equation}}
\newcommand{\ee}{\end{equation}}
\newcommand{\bea}{\begin{eqnarray}}
\newcommand{\eea}{\end{eqnarray}}

\def\1{{\rm 1}}

\def\s1{^{\rm (1)}}

\def\btitle{\begin{Huge}\begin{sffamily}\begin{bfseries}\color{red}}
\def\etitle{\end{bfseries}\end{sffamily}\end{Huge}\normalcolor}
\def\bheader{\begin{LARGE}\begin{sffamily}\begin{bfseries}\color{blue}}
\def\eheader{\end{bfseries}\end{sffamily}\end{LARGE}\normalcolor}

\def\bt{\begin{Huge}\color{red}}
\def\et{\end{Huge}\color{black}}
\def\bh{\begin{LARGE}\begin{bfseries}\color{blue}}
\def\eh{\end{bfseries}\end{LARGE}\normalcolor}
\def\bi{\begin{itemize}}
\def\ei{\end{itemize}}

\title{Compressed Sensing and Reconstruction of Unstructured Mesh Datasets}

\begin{document}

\author{Maher Salloum*, Nathan Fabian, David M. Hensinger, Jeremy A. Templeton}

\maketitle

\bigskip
Sandia National Laboratories, 7011 East Ave., MS 9158,
Livermore,
CA 94550, USA

\bigskip
\begin{center}
mnsallo@sandia.gov
\end{center}


\begin{abstract}
Exascale computing promises quantities of data too large to efficiently store
and transfer across networks in order to be able to analyze and visualize the
results.  We investigate Compressive Sensing (CS) as a way to reduce the size of
the data as it is being stored. CS works by sampling the data on the
computational cluster within an alternative function space such as wavelet
bases, and then reconstructing back to the original space on visualization
platforms. While much work has gone into exploring CS on structured data sets,
such as image data, we investigate its usefulness for point clouds such as
unstructured mesh datasets found in many finite element simulations. We sample
using second generation wavelets (SGW) and reconstruct using the Stagewise
Orthogonal Matching Pursuit (StOMP) algorithm. We analyze the compression ratios
achievable and quality of reconstructed results at each compression rate. We are
able to achieve compression ratios between 10 and 30 on moderate size datasets
with minimal visual deterioration as a result of the lossy compression.

\end{abstract}

\bigskip\bigskip

\section{Introduction}
\label{sec:intro}

Large-scale computing platforms are challenged by the growing size of
computational datasets generated by simulations. On future exascale platforms,
the datasets are expected to be too large to be efficiently stored and
transferred across networks to analysis and visualization workstations, thus
impeding data exploration and knowledge discovery. Several in-situ data
compression schemes are available to reduce the size of simulation datasets. The
compressed version of the dataset is transferred to any workstation and
reconstructed by a scientist for analysis and visualization purposes.

A suitable compression scheme is one that has a small impact on the running
simulation. In other words, the code should not be significantly altered and the
overhead cost should be very low. We propose compressive sensing (CS) as such
method to compress and reconstruct datasets. Starting from the hypothesis that
scientific data has low information density, CS is known to be fast and will
provide a high spatial compression ratio. CS is data agnostic \emph{i.e.} it
does not
require any knowledge of the simulation data type and the features contained
therein, as is the case in wavelet compression. Therefore, it does not require
the selection of a basis type and order during the compression in-situ. The
wavelets bases are selected and computed during the post-processing stage
allowing interactive reconstruction and visualization according to the required
accuracy and quality. Finally, CS is non-intrusive which means that its
implementation does not significantly alter the simulation code.

Conventional CS theory, developed for image compression, is based on the
representation of lattice data using first generation wavelets. There is
currently no literature on the applicability of compressed sensing on
point-cloud data. We will extend the theory to encompass second generation
wavelets (SGW) that can be described on point-clouds \emph{e.g.} an 
unstructured mesh. To our knowledge, this extension of CS to point-cloud data 
has not been explored yet. It will
require designing random matrices that are incoherent with SGW and establishing
the restricted isometry required for unique inflation of compressed samples. 

\subsection{Literature Review}
\label{sec:literature}

In-situ reduction of large computational datasets has recently been the subject
of multiple research efforts. In the ISABELA project \cite{11klasky}, the
dataset is sorted as a vector and encoded with B-splines while computing the
resulting associated errors.  DIRAQ is a more recent effort by the same research
group \cite{lakshminarasimhan2014diraq}. It is a faster parallel in-situ data
indexing machinery that enables an efficient query of the original data. 
ISABELA and DIRAQ suffer from low compression ratios ($\sim 4-6 \times$) which
might not be sufficient for exascale datasets where I/O will be a more critical
bottleneck.  The $\textit{zfp}$ project, \cite{lindstrom2014fixed}, works by
compressing 3D blocks of floating-point field data into a variable-length
bitstream. The $\textit{zfp}$ approach results in compression rates ranging
between one and two orders of magnitude but it is limited to regular grids. 
However, our CS approach compresses unstructured data and we expect to obtain
compression ratios ranging between one and three orders of magnitude.

In-situ visualization \cite{10yu} and feature extraction \cite{13sauer}
are also ongoing research efforts that use combustion simulation datasets
from the S3D simulator developed at Sandia National Labs (SNL). Selected
simulation outputs are analyzed and visualized in-situ.
Data analysis occurs at separate
computational nodes incurring a significant overhead. Moreover, such in-situ
techniques require pre-selected outputs and cannot be used interactively since
most clusters are operated in batch mode. Similar techniques have been
implemented in the Paraview coprocessing library \cite{11fabian}.

Sampling techniques are also employed in-situ for data reduction. Woodring et
al. \cite{11wood} devised such an approach for a large-scale particle
cosmological
simulation. The major feature in this work is the low cost of the offline data
reconstruction and visualization. However, the compression ratios are low and
require skipping levels in the simulation data. Sublinear sampling algorithms
are also proposed for data reduction \cite{13bennett}. Their success has
been proven in the analysis of stationary graphs. They are an ongoing effort at
SNL to transfer sublinear sampling theory into practice with focus on large
combustion datasets.

\bigskip\bigskip

\section{Mathematical Background}
\label{sec:background}

\subsection{Compressive Sensing}
\label{sec:bg_CS}

Compressive Sensing (CS) \cite{08cw2a} was first proposed for image compression
based on the premise that most fields on lattices (images) can be sparsely
represented using first generation wavelets \emph{viz.}, a $N$-pixel image can
be well-approximated with $K$ judiciously chosen wavelets, $K \ll N$. The
non-zero wavelets \emph{i.e.}, the sparsity pattern, are unknown a priori.
Compression can be cast mathematically as the sparse sampling of the dataset in
transform domains. Compressive samples are obtained by projecting the image on
random vectors \cite{06dt2a}, 
\begin{equation}
  \label{eq:samples}
  y = \Phi f, 
\end{equation}

\noindent where, $y \in \mathbb{R}^M$ are the compressed samples, $f \in
\mathbb{R}^N$ is
the field we are sampling, and $\Phi \in \mathbb{R}^{M \times N}$ is the
sampling
matrix. Theoretically, only $C \cdot K \cdot \text{log}_2 (N/K)$ samples are
necessary for reconstruction, where $C$ is a constant that depends on the
data~\cite{08cw2a}.

In practice, reconstruction is posed as a linear inverse problem for the
wavelet coefficients $s \in \mathbb{R}^N$, conditioned on compressive samples
and defined as,
\begin{equation} 
  \label{eq:coefficients}
  f = \Psi s,
\end{equation}

\noindent with $\Psi$ being the wavelet basis matrix. Then the inverse problem
is stated as,
\begin{equation}
  y = \Phi \Psi s = A s,
\end{equation}

\noindent and we must find $s$ from samples $y$ obtained by compression.
Regularization is provided by a
$L_1$ norm of the wavelet coefficients, which also enforces sparsity. The
scalability of the inverse problem solvers (called shrinkage regression) has
only been recently addressed \cite{12borghi}. The random vectors in $\Phi$ are
designed to be incoherent with the chosen wavelets. Incoherence ensures that the
amount of wavelets information carried by the compressive samples is high.
Different types of sampling matrices $\Phi$ can be used to perform the
compression~\cite{06dt2a} (see Eq.~\ref{eq:samples}). In our work, we use the
Bernoulli matrix due to its superior incoherence properties with most basis sets
$\Psi$.

\subsection{Second Generation Wavelets}
\label{sec:bg_wavelets}

Wavelets can represent a basis set that can be linearly combined to represent
multiresolution functions.  Wavelets have compact support and encode all the
scales in the data ($j$) as well as their location ($k$) in space.  There are
two types of wavelets: First generation wavelets (FGW) and second generation
wavelets (SGW), \cite{05janon, 98sweldens}. To date, all CS work of which we
are aware is based on first generation wavelets which only accommodate data
defined on regular grids. FGWs are characterized by scaling functions $\phi(x)$
to perform approximations and $\psi(x)$ to find the details in a function
$f$~\cite{09radunovic}. These are defined at all levels $j$ in the hierarchy and
span all locations $k$ in the regular grid. They are computed in terms of the
so-called mother wavelet $\phi_0$ which provides the approximation at the
largest level $j=0$. At each level $j$, the scaling and detail functions are
defined as,
\begin{equation}
\phi_j (x) = \sum_{k \in \mathbb{Z}} a_k \phi_{j-1} (2x - k),
\end{equation}
\begin{equation}
\psi_j(x) = \sum_{k \in \mathbb{Z}} b_k \phi_{j-1} (2x - k),
\end{equation} 

\noindent where $a_k$ and $b_k$ are constrained to maintain orthogonality. Each
new basis function models a finer resolution detail in the space being spanned.
Regular grids are dyadic and the maximum number of levels $j_\text{max}$ is
equal to log$_2(N)$ where $N$ is the number of grid points in each dimension.

In this work, we examine compression on unstructured data or point clouds which
are not well represented by FGWs~\cite{05janon}. For these unstructured data we
use the SGWs of
Alpert et al.,~\cite{alpert1993wavelet,alpert1993class} referred to as
multi-wavelets. There are two major differences between FGWs and SGWs. The
first difference is that the maximum number of levels $j_\text{max}$, is
computed by recursively splitting the non-dyadic mesh into different
non-overlapping groups, thereby forming a multiscale hierarchy~\cite{05janon}.
Since SGWs do not require dyadic grids, they can accommodate finite intervals
and irregular geometries. The second difference is that in place of
one scaling function, there are several, $\phi_{0,j},\cdots,\phi_{N-1,j}$
defined over groups of the space. In addition, the functions themselves are
defined by the discrete set of points, $x_k$, as opposed to a continuous
representation. The discrete Alpert wavelets $\psi_j(x_k)$ are polynomials, are
quick to compute and are represented in the matrix, $\Psi$. Details on how to
compute the matrix $\Psi$ are given in~\cite{alpert1993wavelet}.

\subsection{Reconstruction}
\label{sec:bg_reconstruction}

Much of the work in CS is handled by the reconstruction phase which uses wavelet
basis to reconstruct the data set from the compressed samples.  Here we use a
greedy algorithm, Stagewise Orthogonal Matching Pursuit (StOMP)~\cite{12dt4a},
that has been empirically demonstrated to be very efficient.

The reconstruction process can be described as follows. We have an
underdetermined linear system,
\begin{equation}
  y = A s,
\end{equation}

\noindent given $y$ and the matrix product $A = \Phi \Psi$, where $\Phi$ is our
sampling matrix and $\Psi$ is the wavelet basis matrix as discussed in
Section~\ref{sec:bg_CS}. We need to find $s$.  If $\Phi$ and $\Psi$ exhibit low
mutual coherence and $s$ is sparse in $\Psi$ \emph{i.e.} has few non-zero
elements, then StOMP can efficiently provide a solution.  StOMP finds the
nonzero elements of $s$ through a series of increasingly correct estimates.
Starting with an initial estimate $s_0 = 0$ and an initial residual $r_0 = y$,
the algorithm maintains estimates of the location of the nonzero elements of
$s$.

StOMP finds residual correlations,
\begin{equation}
  c_n = A^T r_{n-1}, 
\end{equation}

\noindent and uses these to find a new set of nonzero entries, 
\begin{equation}
  J_n = \left\{ j : |c_n(j)| > t_n \sigma_n \right\}.
\end{equation}

\noindent where $\sigma_n$ assumes a Gaussian noise on each entry and $t_n$ is a
threshold parameter we provide in order to assess which coefficients $c_n$ have
to be retained. Elements above the threshold are considered nonzero entries and
added to the set $J_n$.  The new set of entries $J_n$ are added to the current
estimate of nonzero entries $I_n = I_{n-1} \cup J_n$, and used to give a new
approximation of $s_n$ and residual $r_n$,
\begin{equation}
  (s_n)_{I_n} = (A^T_{I_n}A_{I_n})^{-1}A^T_{I_s}y, 
\end{equation}

\noindent and, 
\begin{equation}
  r_n = y - A s_n.
\end{equation}

\section{Implementation}
\label{sec:implementation}

In order to test the procedure, we have built a processing pipeline that both
samples and
reconstructs data from a dataset.  It allows us to experiment with
existing data sets and assess reconstruction quality.  In the
final implementation, the library will be split into two pieces.  The in-situ
piece consists of a small sampling codebase.  It stores the samples, mesh
points and connectivity, and the seed used to generate the Bernoulli sampling
matrix $\Phi$.  During the in-situ processing the sampling matrix is not
constructed explicity, but used implicitly to generate the
sampled data. No wavelet computation is required at this stage.

\begin{figure}[htb]
\centering
\includegraphics[width=0.75\textwidth]{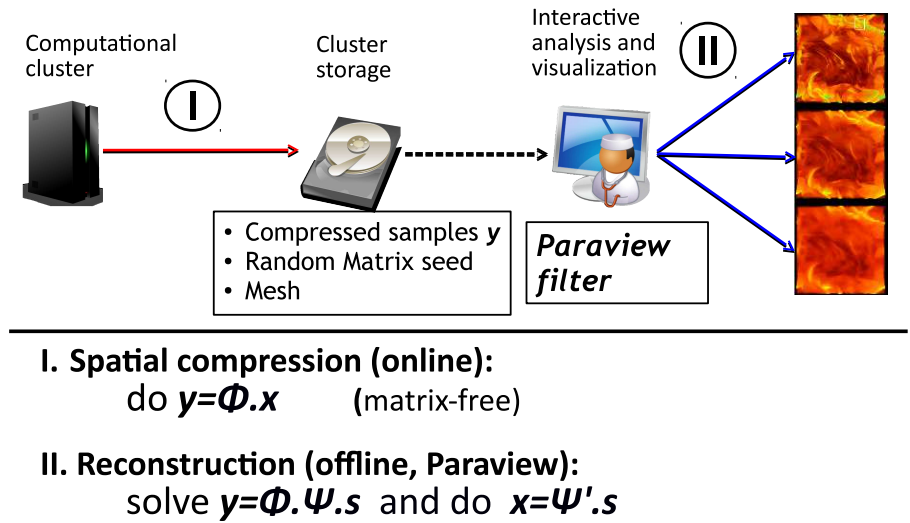}
\caption{\label{fig:schem} A schematic showing the major steps during the 
in-situ data compression using CS (I) and the offline reconstruction and 
visualization of the original dataset (II) using Paraview. The random seed, the
mesh and the compressed samples $y$ are transferred to the 
visualization platform where the reconstruction of $x$ takes place.} 
\end{figure}

The reconstruction side is responsible for rebuilding the sampling matrix and
constructing the wavelet matrix from the mesh data. By providing those matrices
to StOMP, we are able to reconstruct the wavelet sampled data, $s$
in Figure~\ref{fig:schem}, and then inverse transform $s$ to reconstruct
the original data.

We have implemented the reconstruction procedure in two ways: 1) in Matlab in
order to experiment and produce some of the results shown below, and 2) a
more production-oriented version written using the Trilinos sparse linear
algebra library~\cite{trilinos} and ParaView for visualization~\cite{paraview}.

\bigskip\bigskip

\section{Results}
\label{sec:res}

In this section we discuss the compression capability and reconstruction
quality of our method. We also describe some aspects of its practical
implementation. We consider two types of data defined on unstructured meshes.
The first type are ``toy problems" where we assume mathematical functions
defined on an irregular two-dimensional geometry. These datasets are small,
and we can compress and reconstruct them on one processor. The second type are
larger datasets obtained from simulations. In this case, we consider
unstructured three-dimensional meshes distributed among many processors.

\subsection{Two-dimensional datasets}
\label{sec:res2D}
We consider a square geometry with randomly chosen holes such that it
constitutes an irregular geometry. We discretize it using a triangular mesh
consisting of $N=33,067$ nodes. We assume two functions $f$ and $g$ given in
Eqs.~\ref{eq:testhf} and~\ref{eq:testlf} as the datasets represented on the
obtained mesh. We choose these functions such that $f$ exhibits multiple
oscillations and reveals more features than $g$. The motivations behind this
choice is that we would like to explore the effect of data features on the
number of samples $y$ required during compressed sensing, that are necessary to
accurately reconstruct the original dataset.

\begin{align}
f & = 48 \text{sin}(8 \pi x) \text{sin}(7 \pi y) \text{sin}(6 \pi x)
\label{eq:testhf} 
\\
g & = 12 \text{sin}(2 \pi x) \left [ 4 \text{sin}(2 \pi x) - 4
\text{sin}(2 \pi y) \right ]
\label{eq:testlf}
\end{align}

\begin{figure}[htb]
\centering
\includegraphics[width=0.75\textwidth]{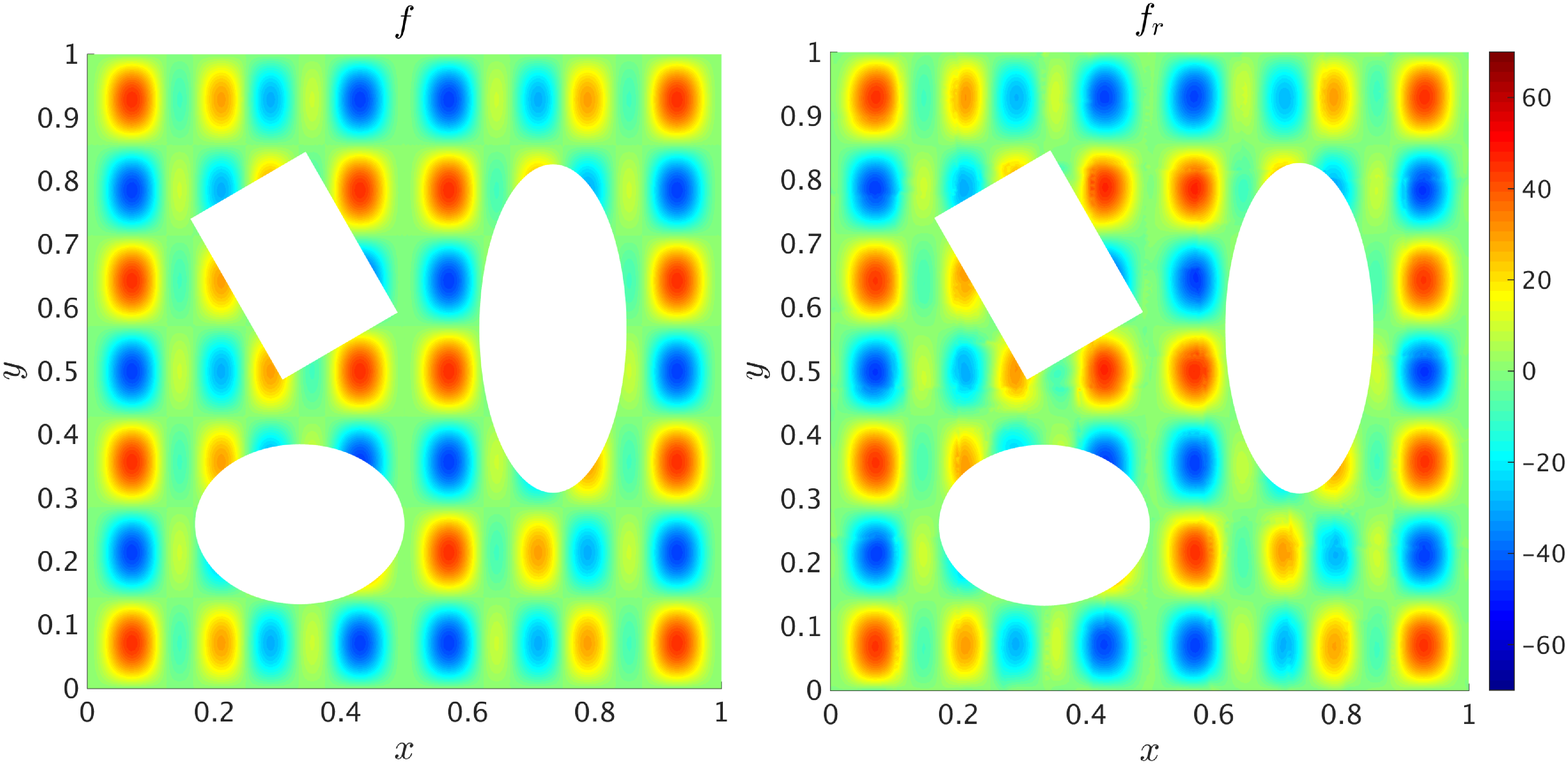}
\bigskip\bigskip
\includegraphics[width=0.75\textwidth]{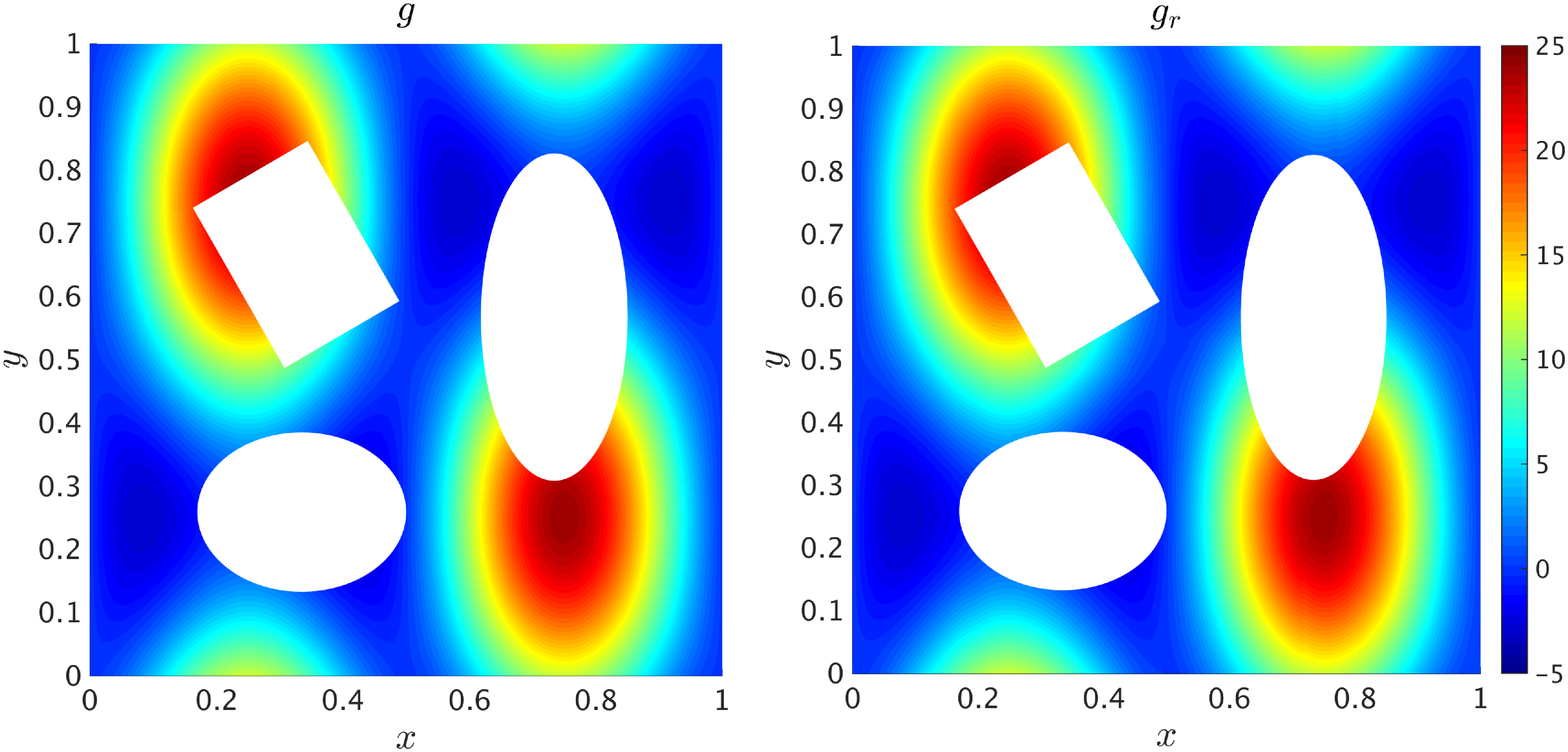}
\caption{\label{fig:hf_R10} Plots showing (left column) smooth datasets
represented on a complex geometry and an unstructured mesh of 33,067 nodes, and
(right column) their reconstructed version from compressed samples, at full
wavelet detail. The top and bottom rows denote two different datasets given by
Eqs.~\ref{eq:testhf} and~\ref{eq:testlf}, respectively. The compression ratios
are $R=10$ and $R=30$ for $f$ and $g$, respectively. The bases for
reconstruction are Alpert wavelets with an order $w=5$.} 
\end{figure}

We compress $f \in \mathbb{R}^N$ and $g \in \mathbb{R}^N$ using a random
Bernoulli matrix $\Phi \in \mathbb{R}^{M \times N}$ as described in
Section~\ref{sec:bg_CS}.
We select an Alpert wavelet basis and reconstruct $f$ using the StOMP algorithm
described in Section~\ref{sec:bg_reconstruction}. Both compression and
reconstruction are performed in a serial run. We denote
the reconstructed fields by $f^r$ and $g^r$, respectively. They are plotted in
Figure~\ref{fig:hf_R10} which shows that the original and reconstructed datasets
are visually identical. The function $f$ that reveals many features; a minimum 
compression ratio of $R=10$ was required for its accurate reconstruction.
However, $g$ affords a higher
compression ratio $R=30$ since it has fewer features which incur more data
redundancy. This result indicates the possibility to predict the compression
ratio in terms of the global gradient. This latter increases with the
number of features in a given dataset. Such prediction of the compression ratio
is extremely important in-situ since the compressibility of the datasets is
unknown a priori.

\begin{figure}[tbh]
\centering
\includegraphics[width=0.372\textwidth]{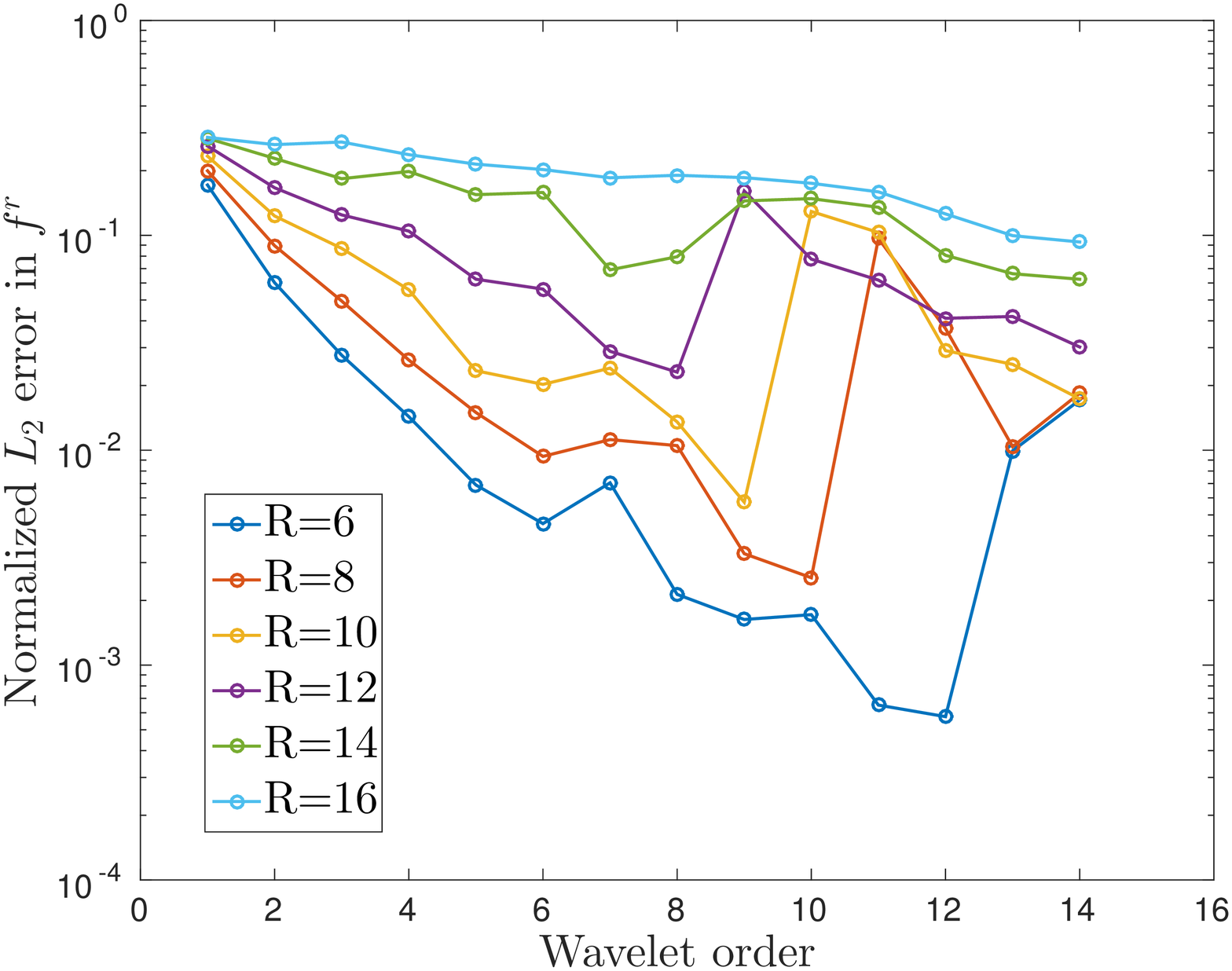}
\includegraphics[width=0.367\textwidth]{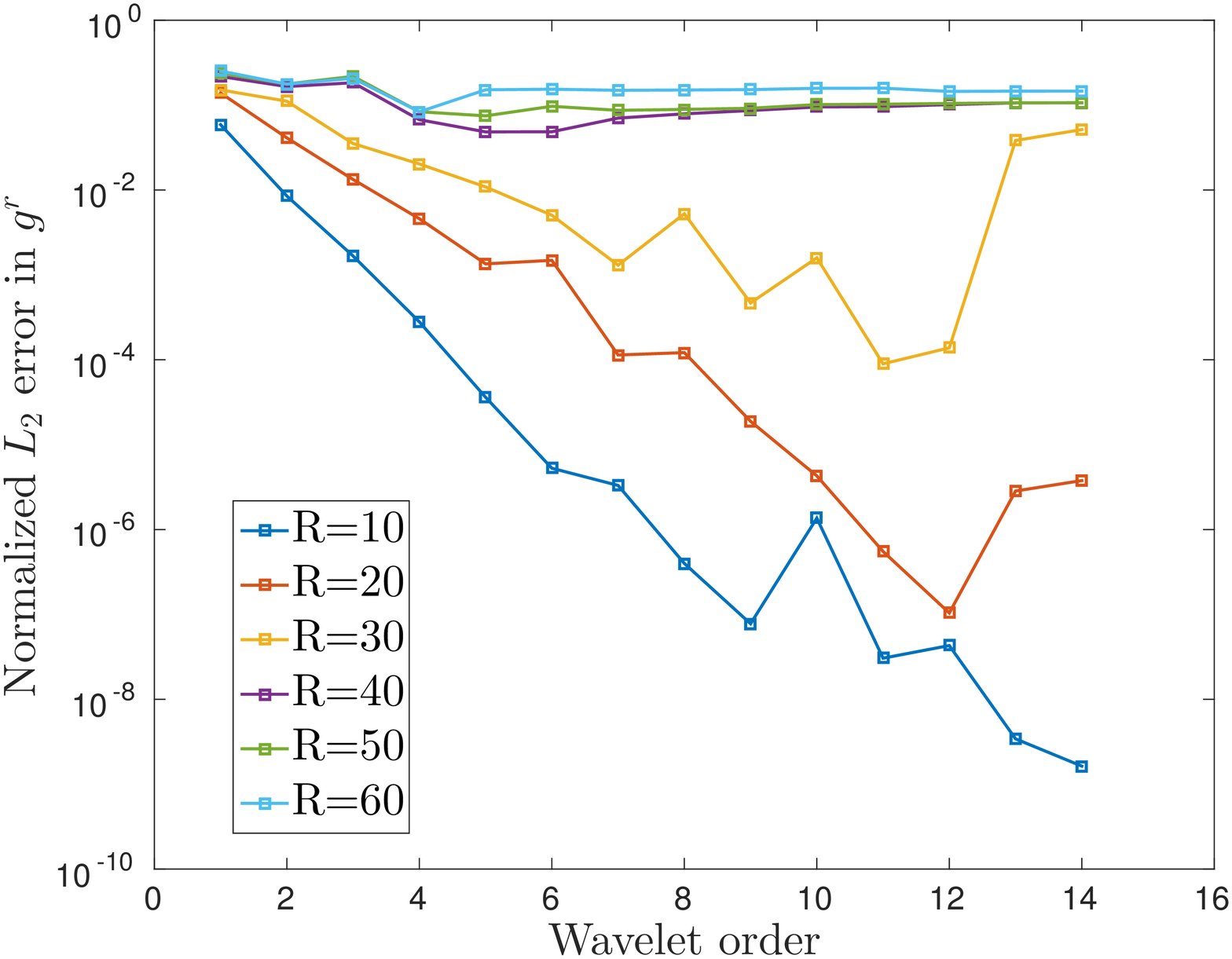}
\caption{\label{fig:errors} Plots showing the normalized $L_2$ error between the
original and reconstructed datasets $f^r$ (left) and $g^r$ (right) plotted in
Figure~\ref{fig:hf_R10} as a function of the wavelet order for different
compression ratios, as indicated.} 
\end{figure}

In order to obtain a quantitative description of the reconstruction accuracy,
we evaluate the global $L_2$ norm of difference between the original and
reconstructed versions. Figure~\ref{fig:errors} shows this error for $f$ and
$g$ as a function of the wavelet order for different compression ratios. As
expected, the error is lower for low compression ratios. We notice that there
exists a minimum compression ratio required to obtain a low reconstruction
error \emph{i.e.} a correct reconstruction. For example, the compression ratio
should be smaller than $30$ for the dataset $g$. This is consistent with the
Donoho chart~\cite{12dt4a} which depicts a discontinuity in the compression
ratio range that guarantees an accurate reconstruction.

The error decreases with the wavelets order $w$. The decreasing trend is
reversed at larger values of $w$. This is mainly attributed to the over-fitting
of the function $f$ by the high order Alpert polynomial wavelets. It is
therefore preferred to choose lower orders. According to the plots in
Figure~\ref{fig:errors}, $w=5$ is an optimum value for the wavelet order that
minimizes the reconstruction error and prevents over-fitting of the given
function in the dataset.

\begin{figure}[tbh]
\centering
\includegraphics[width=0.36\textwidth]{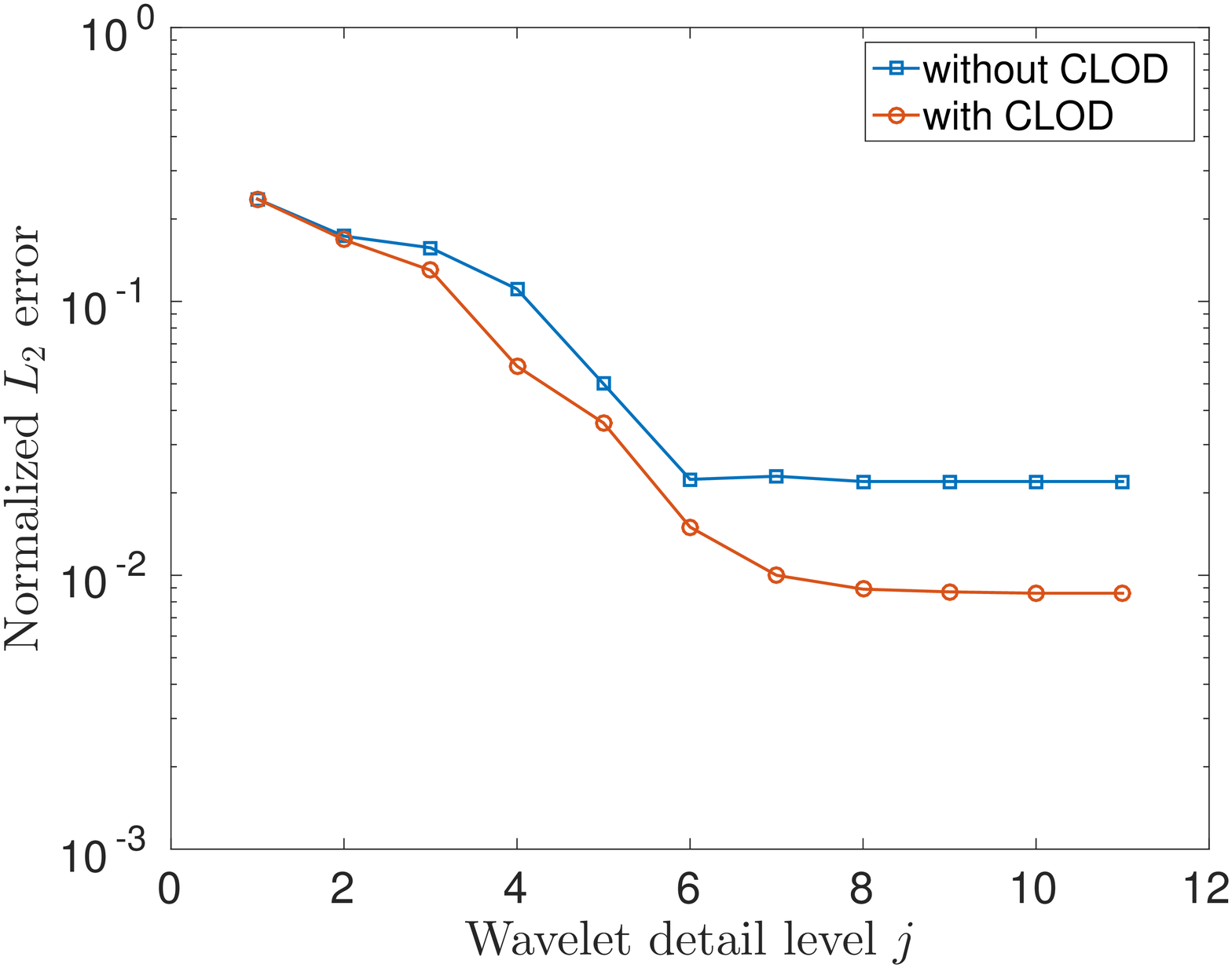}
\includegraphics[width=0.36\textwidth]{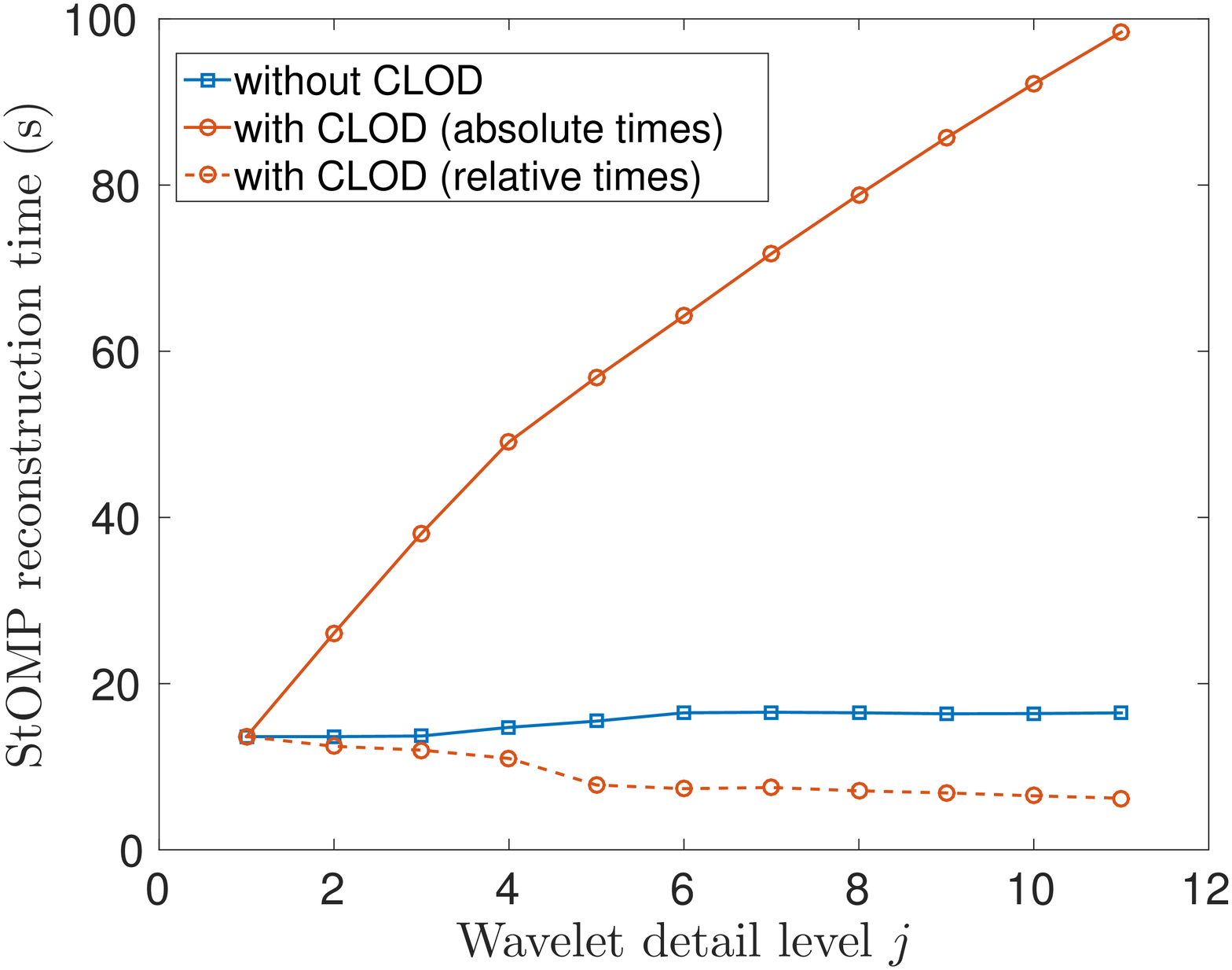}
\caption{\label{fig:errors_times_hf} Plots showing (left) the normalized $L_2$
error and (right) the time taken in a serial run to perform the reconstruction
of the dataset, as a function of the wavelet detail level $j$ for $w=5$. These
results are reported using with and without the CLOD approach, as indicated.} 
\end{figure}

The results presented so far are reported at the full wavelets detail level. We
turn our attention now to the effect of the detail level on the reconstructed
function quality. Such analysis is useful for datasets that exhibit multiple
features. By construction, wavelets are able to represent all scales in a
function where a scale $j$ reveals a level of detail as described in
Section~\ref{sec:bg_wavelets}. In this work, we find the number of detail levels
by performing orthogonal splits along each axis~\cite{98sweldens} which results
in $j_\text{max}=11$ for $w=5$ for the mesh representing $f$. The wavelet matrix
$\Psi$ is computed as the product of different detail level sparse matrices
$\Psi_j$~\cite{alpert1993wavelet,alpert1993class} following:
\be
\Psi = \prod_{j=1}^{j=j_\text{max}} \Psi_j
\label{eq:psi}
\ee

\begin{figure}[htb]
\centering
\includegraphics[width=0.75\textwidth]{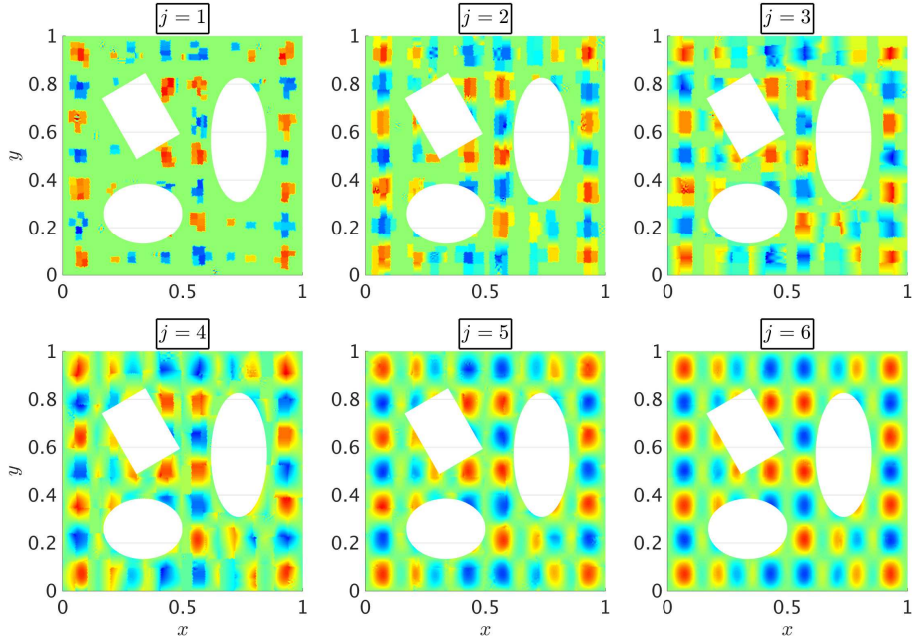}
\caption{\label{fig:plots_CS_levels_hf} Plots showing the dataset in 
Figure~\ref{fig:hf_R10} reconstructed at different wavelet detail levels $j$, 
as indicated. Results are generated from samples compressed with $R=10$ that 
are reconstructed with an Alpert wavelet basis of order $w=5$.} 
\end{figure}

$\Psi$ can be computed at any $1 \le j \le j_\text{max}$ and used to perform a
StOMP reconstruction. The sparsity of $\Psi$ decreases with $j$ to encode more
details. Figure~\ref{fig:plots_CS_levels_hf} shows the reconstructed function
$f^r$ at different detail levels $j$. By scanning the figure, we notice how the
fine details in the function are revealed. Changes between functions
reconstructed at two consecutive detail levels decrease with $j$, indicating
convergence. At $j=6$, the function is visually identical to the original $f$
in Figure~\ref{fig:hf_R10}.

For each detail level, we consider two reconstruction approaches. In the first
approach, the initial guess of the StOMP reconstruction is the same and assumes
wavelet modes equal zero. In the second approach, an initial guess at a detail
level $j$ is obtained from the solution at level $j-1$. Thus, we call this
approach a continuous level of detail (CLOD) reconstruction.
Figure~\ref{fig:errors_times_hf} (left) shows the reconstruction error of the
dataset $f$ as a function of the wavelet detail level. For both approaches, the
error decreases with the detail level, as expected. However, the CLOD approach
results in an error about three times lower at the finest details. This is due
to the accumulation of knowledge in the reconstructed $f^r$ with consecutive
detail levels. Figure~\ref{fig:errors_times_hf} (right) shows the reconstruction
time required at each level. Using CLOD, the relative time taken between two
consecutive levels decreases with $j$ due to updated initial guess resulting in
a decreased number of StOMP iterations. However, the cumulative time
is substantially higher than the case without CLOD. The additional time required
by CLOD to reach a smaller reconstruction error can be alleviated by skipping
the reconstruction at some levels \emph{e.g.} performing the CLOD reconstruction
at the odd numbers levels.

\subsection{Three-dimensional distributed datasets}
\label{sec:res3D}

\begin{figure}[htb]
\centering
\includegraphics[width=0.75\textwidth]{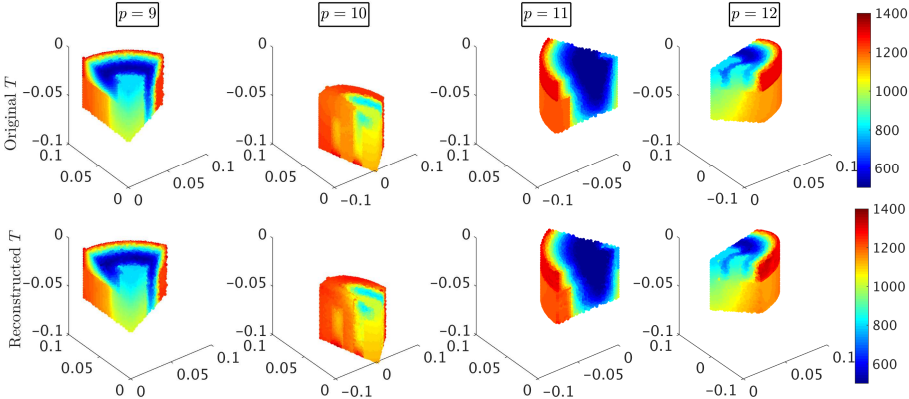}
\caption{\label{fig:maze3d} Plots showing the steady state temperature
distributions in
different portions of the three-dimensional (top) original and (bottom)
reconstructed dataset $T$. Results are reported from a parallel simulation on
16 processors where compression is performed locally with a ratio $R=10$. The
vertical panels correspond to almost equal portions of the full dataset as
distributed among different processors, as indicated. The reconstruction is
performed using StOMP for a wavelet order $w=5$ at full detail.} 
\end{figure}

In this section, we consider a larger dataset represented on a three-dimensional
tetrahedral mesh with 396,264 points. The data is the temperature field
obtained by a transient heat conduction simulation. The cylindrical geometry
constitutes several sub-domains of different solid materials. The sub-domains
sizes, heat conductivity and heat generations rates are chosen in a random
fashion which results in a heterogeneous temperature distribution. Initially, 
the temperature is uniform across the domain, after which, it evolves to a
steady state. The simulation
is performed in parallel on 16 processors and the datasets is split equally
among the processors (~24,766 point per processor). The compression is performed
locally on each processor \emph{i.e.} the matrix-vector product in
Eq.~\ref{eq:samples} was performed serially on each processor with no
communications with other processors. Doing so preserves an efficient in-situ
compression, and a faster and memory-efficient StOMP reconstruction on the
visualization workstation. The whole dataset can be recovered by assembling the
different reconstructed portions. Figure~\ref{fig:maze3d} shows the original and
reconstructed versions of the steady state temperature fields for different
portions of the dataset corresponding to different processors. For a compression
ratio $R=10$, the reconstructed and original datasets are in a visual agreement.

\begin{figure}[htb]
\centering
\includegraphics[width=0.75\textwidth]{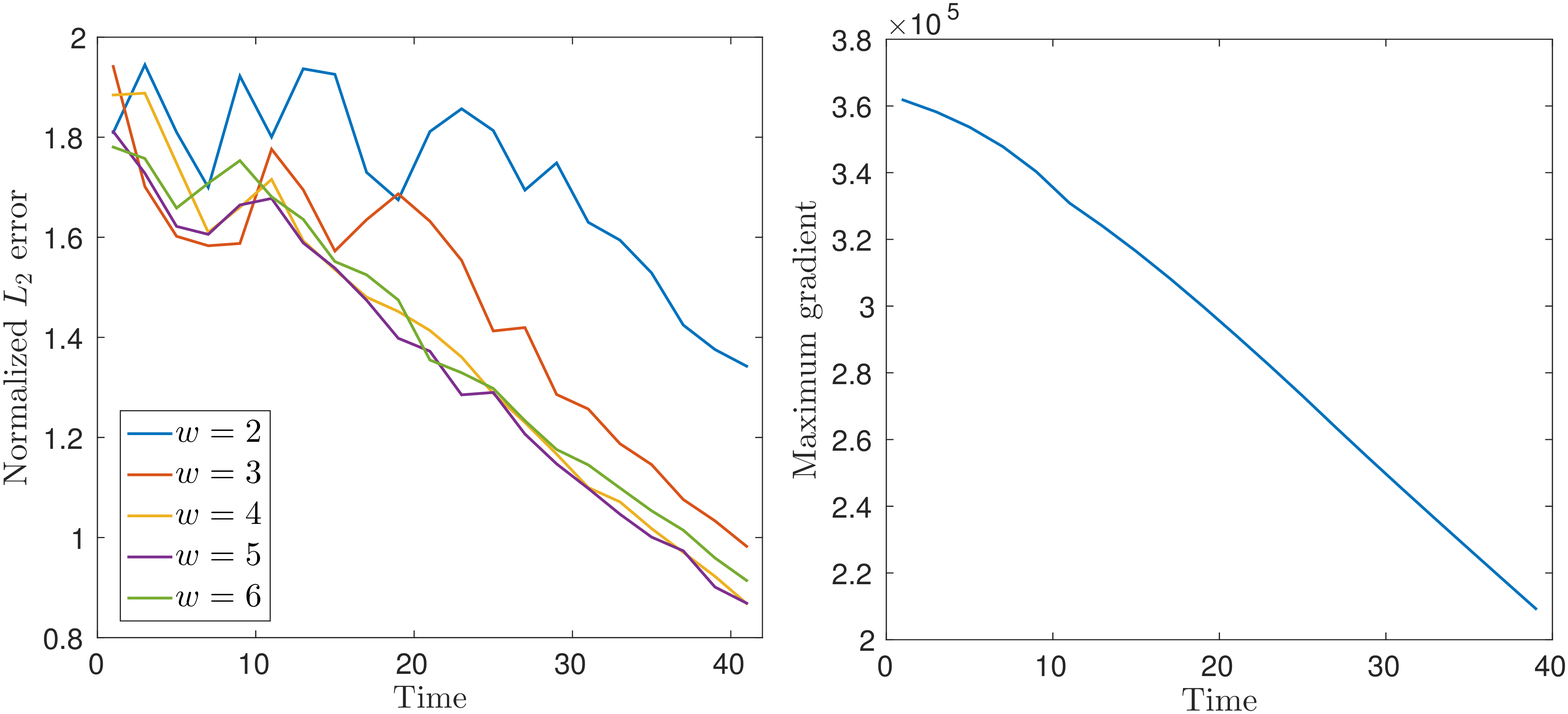}
\caption{\label{fig:maze3d_error} Plots showing the variation of the 
reconstruction error in the dataset $T$ as a function of time. Results are 
reported from a parallel simulation on 16 processors where compression is 
performed locally with a ratio $R=10$. The reconstruction is performed using 
StOMP at full detail for different wavelet orders $w$, as indicated.} 
\end{figure}

Finally, we report in Figure~\ref{fig:maze3d_error} (left) the evolution of the
reconstruction error as a function of time for a constant compression ratio
$R=$.10 The error is initially large since at earlier times, large gradients
exist in the dataset. As heat diffuses in the domain, the temperature field
becomes smoother and better represented by the wavelet bases. It leads to a
smaller reconstruction error. Overall, the error decreases with larger wavelet
orders as expected. We notice a saturation in the error and even a slightly
increased error for higher orders mainly at early times. Similar to the test
cases in Section~\ref{sec:res2D}, this is due by the over-fitting of the high
gradients in the temperature field by the higher order Alpert wavelets. These
results suggest that local large gradients form discontinuities and contribute
to the overall reconstruction error. Therefore, when predicting the compression
ratio in-situ, the local gradients have to be accounted for along the global
gradient discussed in Section~\ref{sec:res2D}.

\bigskip\bigskip

\section{Conclusion}
\label{sec:conc}

In this paper we have demonstrated an application of compressive sensing to
unstructured mesh data. We used second generation wavelets to efficiently
represent the irregularities present in the spatial geometries and meshes. We
are able to achieve lossy compression ratios of between 10 and 30 on fields
defined on these meshes, depending on the oscillations and features present in
the data. The visual deterioration as a result of the lossy compression at those
rates is minimal. Large gradients and discontinuities in the data also
contribute in assessing the reconstruction quality. We explored continuous level
of detail reconstruction for datasets exhibiting many features and found that it
results in a lower reconstruction error at the expense of an increased
computational cost. We continue to explore ways to improve the algorithms used
here in terms of reconstruction time and streaming.  It may also be the case
that other wavelet and sampling matrix pairs and reconstruction algorithms will
produce better results on some data.  We continue to investigate these as well.

\bigskip\bigskip

\section{Acknowledgements}
The authors would like to acknowledge Dr. Jaideep Ray for providing
valuable discussions and feedback that were helpful to accomplish this
work.

\bigskip

This work was supported by the Laboratory Directed Research and Development
(LDRD) program at Sandia National Laboratories.

\bigskip

Sandia National Laboratories is a multi-program laboratory managed and
operated by Sandia Corporation, a wholly owned subsidiary of Lockheed
Martin Corporation, for the U.S. Department of Energy's National
Nuclear Security Administration under contract DE-AC04-94AL85000.
\bigskip\bigskip

\bibliographystyle{unsrt}
\renewcommand{\refname}{\section{References}}
\bibliography{comp_sens}

\end{document}